%% file: main.tex
\documentclass[sigconf]{acmart}

\usepackage{helvet}
\usepackage{courier}
\usepackage{epsfig}
\usepackage{float}
\usepackage{color}
\usepackage{booktabs}
\usepackage{graphicx}
\usepackage{caption}
\usepackage{subcaption}
\usepackage{amsfonts}
\usepackage{enumitem}

\newcommand{\nop}[1]{}
\newcommand{\modelsup}[0]{TER}
\newcommand{\modelsemi}[0]{TER+}

\setcopyright{none}

\nop{
	\acmDOI{10.475/123_4}
	
	\acmISBN{123-4567-24-567/08/06}
	
	\acmConference[WOODSTOCK'97]{ACM Woodstock conference}{July 1997}{El
		Paso, Texas USA} 
	\acmYear{1997}
	\copyrightyear{2016}

	\acmPrice{15.00}
}

\begin{document}

\title{Joint Text Embedding for Personalized Content-based Recommendation}

\author{Ting Chen}
\authornote{Work done while the first author was an intern at Yahoo! Research.}
\affiliation{%
	\institution{University of California, Los Angeles}
	\city{Los Angeles}
	\state{CA}
}
\email{tingchen@cs.ucla.edu}

\author{Liangjie Hong}
\affiliation{%
	\institution{Etsy Inc.}
	\city{New York City}
	\state{NY}
}
\email{lhong@etsy.com}

\author{Yue Shi}
\authornote{Now at Facebook.}
\affiliation{%
	\institution{Yahoo! Research}
	\city{Sunnyvale}
	\state{CA}
}
\email{yueshi@acm.org}

\author{Yizhou Sun}
\affiliation{%
	\institution{University of California, Los Angeles}
	\city{Los Angeles}
	\state{CA}
}
\email{yzsun@cs.ucla.edu}

\nop{

	
}

\input{abstract}
\maketitle

\input{introduction}
\input{problem}

\input{model}
\input{exp}
\input{related}
\input{conclusion}

\section*{Acknowledgements}
The authors would like to thank Qian Zhao, Yue Ning, and Qingyun Wu for helpful discussions. Yizhou Sun is partially supported by NSF CAREER \#1741634.

\bibliographystyle{ACM-Reference-Format}
\bibliography{main}

\end{document}

%% file: abstract.tex
\begin{abstract}

Learning a good representation of text is key to many recommendation applications. Examples include news recommendation where texts to be recommended are constantly published everyday. However, most existing recommendation techniques, such as matrix factorization based methods, mainly rely on interaction histories to learn representations of items. While latent factors of items can be learned effectively from user interaction data, in many cases, such data is not available, especially for newly emerged items.

In this work, we aim to address the problem of personalized recommendation for completely new items with text information available. We cast the problem as a personalized text ranking problem and propose a general framework that combines text embedding with personalized recommendation. Users and textual content are embedded into latent feature space. The text embedding function can be learned end-to-end by predicting user interactions with items. To alleviate sparsity in interaction data, and leverage large amount of text data with little or no user interactions, we further propose a joint text embedding model that incorporates unsupervised text embedding with a combination module. Experimental results show that our model can significantly improve the effectiveness of recommendation systems on real-world datasets.
\end{abstract}

%% file: introduction.tex
\section{Introduction}

Personalized recommendation has gained a lot of attention during the past few years~\cite{koren2009matrix,salakhutdinov2007restricted,wang2011collaborative}. Many models and algorithms have been proposed for personalized recommendation, among which, collaborative filtering techniques such as matrix factorization~\cite{salakhutdinov2011probabilistic,koren2008factorization} are shown to be most effective. For these approaches, historical behavior data is critical for learning latent factors of both users and items. However, in many scenarios, behavior data is not available or very sparse, which motivates us to incorporate content/text information for recommendation. In this work, we study the problem of content-based recommendation for \textit{completely new} items/texts, where historical user behavior data is not available for new items at the time of recommendation. However, when it comes to text article recommendations, it is not straightforward to incorporate text content into existing collaborative filtering models. 

In order to understand content of new items/texts for better recommendation, a good representation based on textual information is essential. This issue is challenging and has not been satisfyingly solved yet. On one hand, traditional content-based~\cite{pazzani2007content} recommendation methods are usually based on simple text processing methods such as cosine similarity or logistic regression where both text and users are represented as bag-of-words. The limitations of such representation include the inability to encode similarity between words, as well as losing word order information~\cite{mikolov2013distributed,johnson2014effective}. On the other hand, for collaborative filtering methods, although some of which has been extended to incorporate auxiliary information, text feature extraction functions are usually simple, and cannot leverage recent proposed representation learning techniques for text~\cite{singh2008relational,rendle2010factorization,chen2012svdfeature}.

We address these issues with an approach that marries text embedding to personalized recommendation. In our proposed model, users and texts are simultaneously embedded into latent space where preferences can be depicted by simple dot product. While each user is directly associated with an embedding vector, text embedding requires an embedding function that maps a text sequence into a vector. Both user embedding and text embedding function can be trained end-to-end based on user-item interactions directly. With sophisticated neural networks (e.g., Convolutional Neural Networks) as text embedding function, high-level textual features can be better captured.

While end-to-end training of the embedding function delivers focused supervision for learning the task related representations. Interaction data is usually sparse, and there are still large amount of unlabeled data/corpora. Hence, we further propose a joint text embedding model to leverage unsupervised text embeddings that are pre-trained on large-scale unlabeled copora. To effectively fuse both types of information, a novel combination module is constructed and incorporated into the unified framework. Experimental results on two real-world data sets demonstrate the effectiveness of the proposed joint text embedding framework.

%% file: problem.tex
\section{Preliminaries and Related Work}

\subsection{Problem Definition}

We use $X = (x_1, \cdots, x_N)$ to denote the set of texts, the $i$-th text is represented by a sequence of words, i.e. $x_i = (w_1, \cdots, w_t)$. A matrix $C$ is used to denote the historical interactions between users and texts, where $C_{ij}$ indicates interaction between a user $i$ and a text article $j$, such as click-or-not, like-or-not. \footnote{We consider $C$ as implicit feedback in this work, which means only positive interactions are provided, and non-interactions are treated as negative feedback implicitly.}

Given text information $X$ and historical interaction data $C$, our goal is to learn a model which can rank completely new texts for an existing user $i$ based on this user's interests and the text content.

\subsection{Personalized Recommendation}

Existing methods of personalized recommendation algorithms can be roughly categorized into there categories: (1) collaborative filtering methods, (2) content-based methods, and (3) hybrid methods.

Matrix factorization (MF) techniques~\cite{koren2008factorization,salakhutdinov2011probabilistic} is one of the most effective collaborative filtering (CF) methods. In MF, each user or item is associated with latent factor vectors $u$ or $v$ and the score between a pair of user and item is computed by their dot product, i.e. $s_{ij} = u_i^T v_j$. Since each item $j$ is associated with latent factors $v_j$, an new item cannot be handled properly as the training of $v_j$ depends on its interaction with users. 

Content based methods~\cite{pazzani2007content} usually build model for user and content based on term weighting schemes like TF-IDF. And cosine similarity or logistic regression can be used to match between a pair of user and item. It is difficult to work with such representations to encode similarities between words, as well as word orders.

Hybrid methods can improve so-called ``cold-start" issue by incorporating side information~\cite{rendle2010factorization,chen2012svdfeature,singh2008relational}, or item content information~\cite{wang2011collaborative,gopalan2014content,wang2015collaborative}. However, most of these methods cannot deal with completely new items. 

There are some work aiming at leveraging neural networks for better text recommendations, such as Collaborative Deep Learning~\cite{wang2015collaborative}, and others~\cite{bansal2016ask}. Compared to their work, 1) we treat the problem as a ranking problem instead of a rating prediction problem, thus pairwise loss functions are adopted; 2) our model provide a more general framework, enabling various text embedding functions, thus subsumes \cite{bansal2016ask} as a special case; 3) our model incorporates unsupervised text embedding from large-scale unlabeled corpora.

\subsection{Text Embedding}

Recent advances in deep learning have demonstrated the importance of learning good representations for text and other types of data~\cite{mikolov2013distributed,mikolov2013efficient,le2014distributed,kim2014convolutional,chen2016entity,chen2017task}. Text embedding techniques aim at mapping text into vector representation that can be utilized for future predictive tasks. Such models have been proposed for addressing text classification/categorization problem~\cite{kim2014convolutional,johnson2014effective,le2014distributed}. Our task resembles a personalized text classification/ranking problem, in the sense that we try to classify/rank an article according to its interestingness w.r.t. a given user. Also, we utilize user behavior instead of labels of text as a supervised signal.

%% file: model.tex
\section{The Proposed Model}

In this section, we first introduce the supervised text embedding framework, which is trained in an end-to-end fashion for predicting user-item interactions. Then we propose a joint text embedding model by incorporating unsupervised text embedding with a combination function.

\subsection{Supervised Text Embedding Framework}

To simultaneously capture interests of users and semantics of texts, we embed both user and text into a common latent feature space, where dot product can be used to quantify their proximity.

Each user $i$ is directly associated with an embedding vector $u_i$, which represents user's interests. For a text sequence $x_j$ for the $j$-th item, it is mapped into a fixed-sized vector by an embedding function $f(x_j) \in \mathbb{R}^k$. The proximity score $s_{ij}$ between the user and item pair $(i, j)$ is computed by the dot product between their embeddings, as follows: 
$$
s_{ij} = u^T_i f(x_j)
$$
\textbf{Text embedding function $f(x)$}. In our framework, the text embedding function is very flexible. It can be specified by \textit{any} differentiable function that maps a text sequence into a fix-sized embedding vector. Many neural network structures can be applied, such as Convolutional Neural Networks, Recurrent Neural Networks, and etc. Here we introduce two such functions, \textit{MoV} and \textit{CNN}, while other extensions are straightforward.

\begin{figure}
	\centering
	\includegraphics[width=0.45\textwidth]{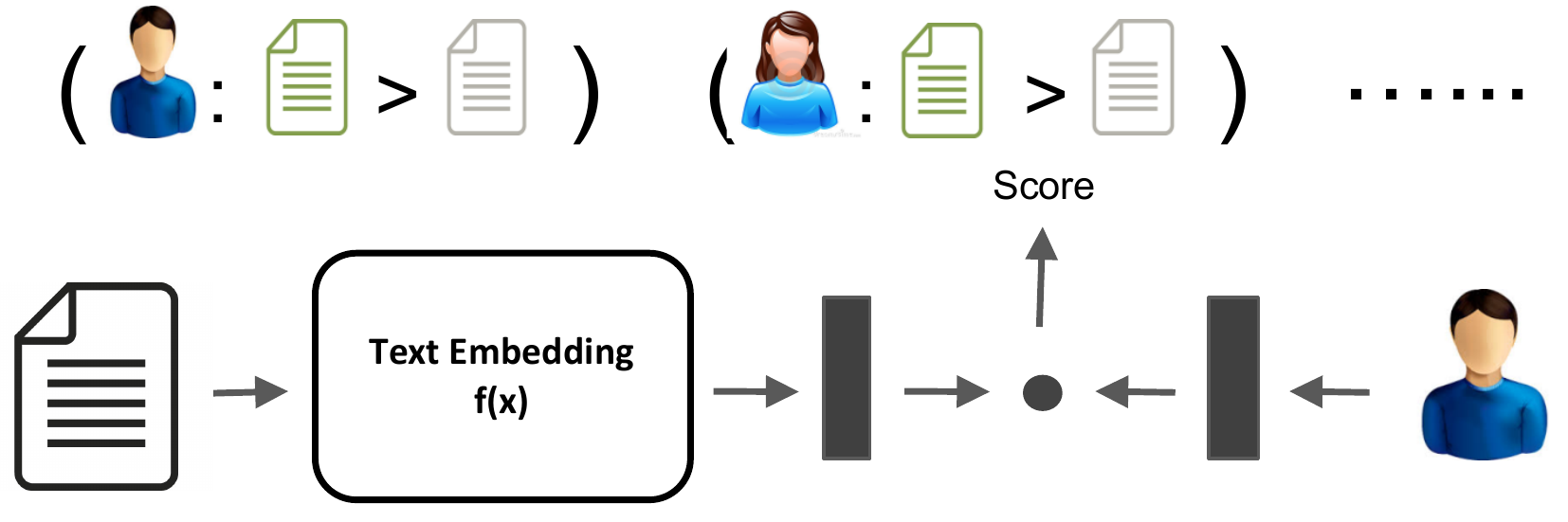}
	\caption{A supervised text embedding framework. Predicted score for a user-text interaction is fit into pairwise ranking objective shown on the top.}
	\label{fig:arch_sup}
\end{figure}

\begin{figure*}[t!]
	\centering
	\begin{subfigure}[b]{0.5\textwidth}
		\includegraphics[width=\textwidth]{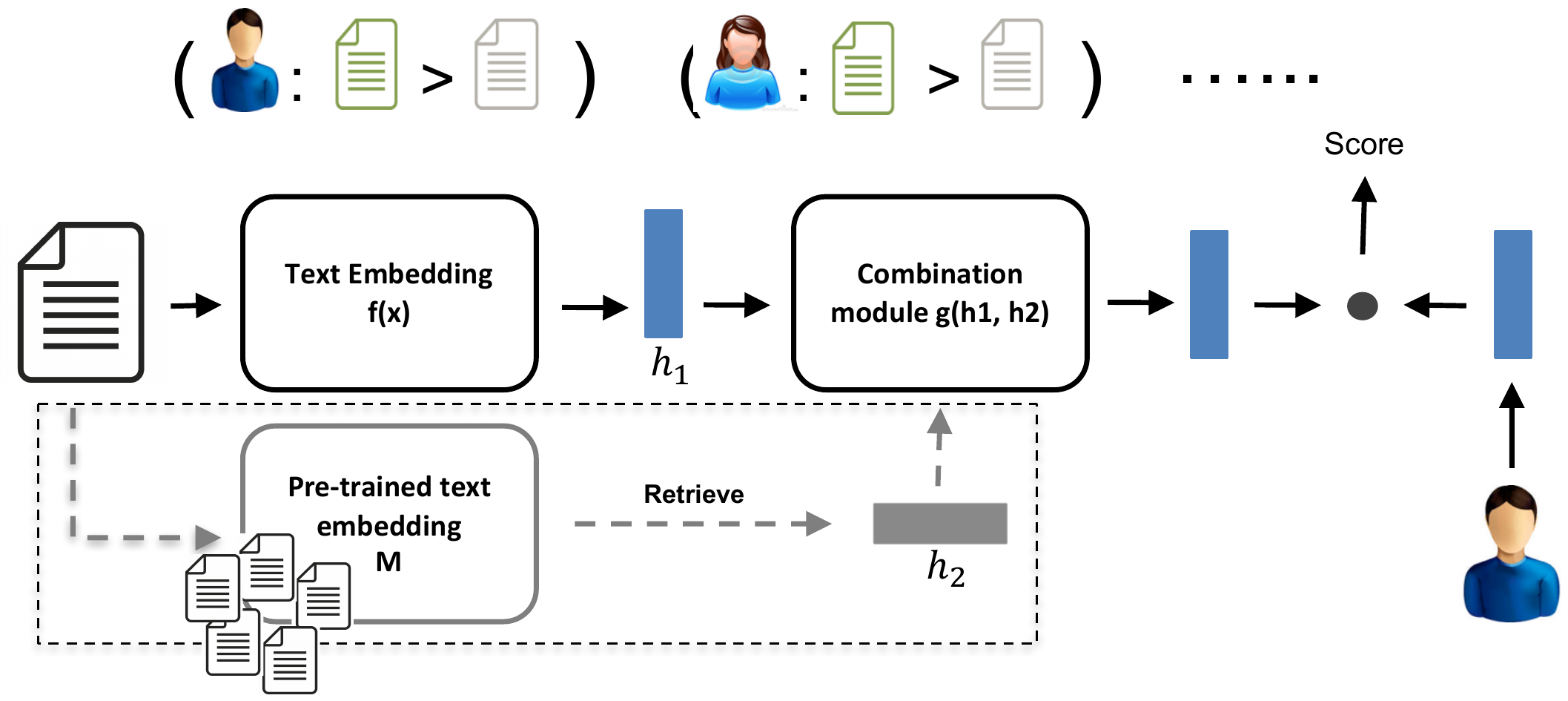}
		\caption{The extended text embedding framework. The unsupervised text embedding is first trained based on a unlabeled text corpus. Then the remaining is trained according to personalized ranking of items.}
		\label{fig:arch_semi}
	\end{subfigure}
	\hspace{4em}
	\begin{subfigure}[b]{0.22\textwidth}
		\includegraphics[width=\textwidth]{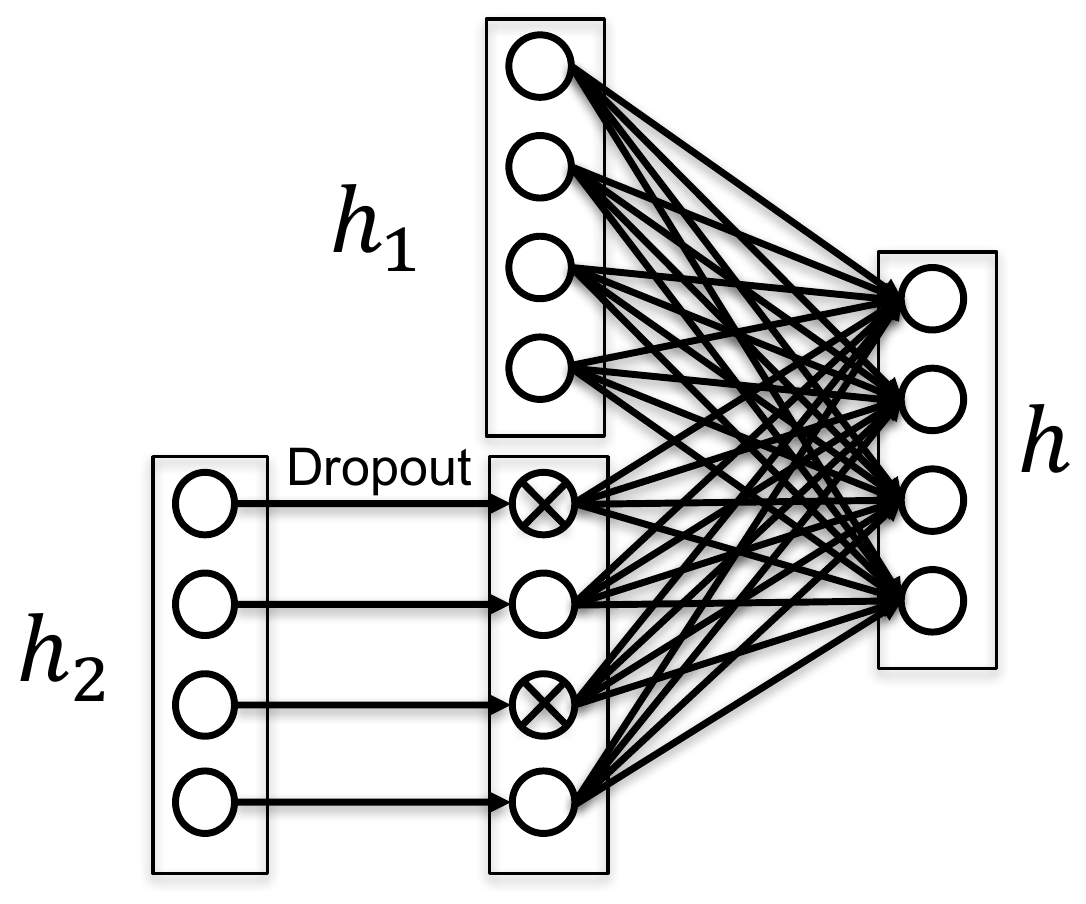}
		\caption{Combination function $g(h_1, h_2)$ explicitly combines two text embedding vectors.}
		\label{fig:arch_comb}
	\end{subfigure}
	\caption{The joint text embeding framework.}\label{fig:arch_semi_more}
\end{figure*}

\textit{Mean of Vectors (MoV)}. To represent a text sequence $x$ of length $T$, we first embed each word in the text with an embedding vector $w$~\cite{mikolov2013distributed,mikolov2013efficient}, and then use the average of word embeddings to form the text/article embedding as follows:
$$h = \frac{1}{t} \sum_{i=1}^{T} w_{x_i}$$
To better extract non-linear interactions among words, a densely-connected layer with non-linear activation can be applied. A single layer of such transformation is given by:
$$
h' = \mbox{Relu}(W h + b)
$$
where $\mbox{Relu}(\cdot) = \mbox{max}(0, \cdot)$ is Rectified Linear Unit.

\textit{Convolutional Neural Networks (CNN)}. Although \textit{MoW} model is simple and relatively efficient, since text sequence is treated as a bag of words, orderings among words are ignored. As demonstrated in~\cite{johnson2014effective},  ordering information of words can be helpful. To address the issue, Convolutional Neural Networks is adopted for text embedding. 

In CNN, instead of averaging over all word embeddings, it maintains several filters of given size(s). Each filter will slide over the whole text sequence. Additionally, at each position, an activation is computed by a dot product between the filter and local embedding vectors. To be more specific, we use $D = w_{x_1}\oplus w_{x_2}\oplus\cdots \oplus w_{x_T} \in \mathbb{R}^{T \times k}$ to denote the concatenation of word vectors for the text. To apply convolution on text sequence $D$, we compute the $j$-th entry from applying $i$-th filter according to:
$$
c^i_j = \mbox{Relu}(W^j \cdot D_{i:i+s-1} + b^j). 
$$
Here $W^j\in \mathbb{R}^{s\times k}$ is the $j$-th filter of size $s$, and $b^j$ is the bias term. The output $c$ of convolution layer can be downsized by pooling operator, such as taking max over all temporal dimensions of c, so a fixed sized vector $h$ can be produced. Due to the page limit, we refer the reader to~\cite{kim2014convolutional} for more clear detailed descriptions.

\textbf{Objective Function and Training}. To learn the user embedding and text embedding function, the output scores for each pair of user and item are used to predict their interactions. For a given user, we want to rank his/her interested articles higher than those he/she is not. So for each user $i$, a pair of a positive item $p$ and a negative item $n$ are both sampled, and similar to~\cite{rendle2009bpr}, the score difference between positive and negative items is maximized, leading to a pairwise ranking loss function as follows:
\begin{equation}
\label{eq:objective}
\mathcal{L} = \mathbb{E}_{(i, p, n) \sim \mathcal{D}}\bigg[ -\log \sigma ( s_{ip} - s_{in})\bigg]
\end{equation}
where $p$ is a positive item for user $i$, and $n$ is a negative item for user $i$. Each triplet $(i, p, n)$ is drawn from some predefined data distribution $\mathcal{D}$. And $\sigma$ is sigmoid function. The objective can be optimized by Stochastic Gradient Descent (SGD). In order to get triplets $\{(i, p, n)\}$ for training, positive interactions $\{(i, p)\}$ are first sampled, and then negative items are sampled according to some predefined distribution (e.g. item frequency).

The framework is demonstrated in Figure~\ref{fig:arch_sup}. We name the above proposed framework \modelsup{}, short for \underline{T}ext \underline{E}mbedding for content-based \underline{R}commendation.

\subsection{Incorporating Unsupervised Text Embedding}

There are two challenges faced by the supervised text embedding framework proposed above: 1) user-item interaction data may be sparse, and 2) there are many texts with little to none user interactions. These issues can lead to over-fitting. To alleviate sparsity in interaction data and leverage a large amount of text data with little to none user interactions, we propose to incorporate unsupervised text embedding with a new combination function. The overall framework with joint text embedding is summarized in Figure~\ref{fig:arch_semi}.

Different from the supervised model, a pre-trained text embedding module is added, so each text is first mapped into two embedding vectors: $h_1$ from text embedding function $f(x)$ and $h_2$ from pre-trained embedding matrix $M$. Then to generate a cohesive text embedding vector for the item, we propose a combination function $g(h_1, h_2)$ to explicitly combine $h_1$ and $h_2$. Below we introduce these two additional components in detail.

\textbf{Unsupervised Text Embedding Matrix $M$}. Unlike supervised text embedding, which requires user interactions for training mapping function $f(x)$. The unsupervised text embedding can be pre-trained with only text articles themselves, requiring no additional labels. To leverage a large-scale text corpus, we adopt Paragraph Vector~\cite{le2014distributed} in our framework. 

Given a set of text articles, Paragraph Vector associates each word $i$ with a word embedding vector $w_i$ and each document $d$ with a document embedding vector $v_d$. To learn both types of embedding vectors simultaneously, a prediction task is formed: for each word occurrence, we firstly hide the word, and then model is asked to predict the exact word given neighboring word embeddings and the document embedding. The probability of the word is given as follows:
$$
P(token = i) = \frac{\exp(v_d^T w_i)}{\exp(\sum_j v_d^T w_j)}
$$As introduced in \cite{le2014distributed}, the model is trained by maximum likelihood with negative sampling.

After training Paragraph Vector on the whole corpus, which includes text articles that have no related user interaction associated. We obtain a pre-trained text embedding module with embedding matrix $M$, where each row $M_i$ is an unsupervised text embedding vector for the $i$-th text article.

\textbf{Combination Function $g(h_1, h_2)$}. To combine both text embedding vectors, i.e. $h_1$ from text embedding function $f(x)$, and $h_2$ from pre-trained embedding matrix $M$, we introduce a combination function $g(h_1, h_2) \in \mathbb{R}^k$ where $k$ is the user-defined output dimension. Since the relation between two text embedding vectors $h_1$ and $h_2$ can be complicated and non-linear, in order to combine them effectively, we specify the combination function $g(\cdot)$ with a small neural network: Firstly a concatenation of the two vectors are formed, i.e. $h = [h_1, h_2]$, and then it is further transformed by a densely-connected layer with non-linear activations, i.e. $h' = \mbox{Relu}(Wh + b)$.

Although unsupervised text embeddings can provide useful text features~\cite{le2014distributed}, they might not be directly relevant to the task. So to control the degree of trust for unsupervised text embeddings, we introduce dropout~\cite{srivastava2014dropout} into unsupervised text vectors, i.e. $h_2$, which randomly select entries and set them to zero. On one hand, when setting the dropout to zero, the whole embedding vector is utilized; on the other hand, when setting the dropout to one, the whole text vector is set to zero, hence it is equivalent to use none of pre-trained embeddings. When the dropout rate is between zero and one, it can be seen as a trade-off for the unsupervised module. Figure~\ref{fig:arch_comb} illustrates the combination module.

\textbf{Training of the Joint Model}. The training procedure is separated into two stages. At the first stage, a unsupervised text embedding matrix $M$ is trained using unlabeled texts. At the second stage, similar to the supervised framework, the training objective is also pairwise ranking objective in Eq.~\ref{eq:objective}. The parameters in the second stage involve both user embeddings and parameters in $f(x)$ and $g(h_1, h_2)$. Finally we name the extended model \modelsemi{}.

%% file: exp.tex
\section{Experiments}
In this section, we present our empirical studies on two real-world text recommendation data sets.

\subsection{Data Collections}
Two real-world data sets are used. The first data set CiteULike, containing user-bookmarking-article behavior data from CiteULike.org, was provided in~\cite{wang2011collaborative}. It contains 5,551 users, 16,980 items, and 204,986 interactions. The second data set is Yahoo! News Feed\footnote{https://webscope.sandbox.yahoo.com/catalog.php?datatype=r\&did=75}. We randomly sampled 10,000 users (with at least 10 click behaviors) and their clicked news to form the data set, which contains 58,579 items, and 515,503 interactions. Since CiteULike and News data sets have both title and abstract/summary, for each data set, we create following two data sets: one contains only title information (i.e. short text), and the other contains both title and summary/abstract (i.e. long text). The average lengths of short text in CiteULike and News are 9 and 11 respectively, and that of long text are 194 and 89 respectively.

To ensure items at the test time are \textit{completely new}, we first select a portion (20\%) of items to form the pool of test items. All user interactions with those test items are held-out during training, only the remaining user-item interactions are used as training data. \nop{Note that users that are only appeared in test records are excluded in test data.} For unsupervised text embedding pre-training, we also include many texts that have no user interaction data. More specifically, for CiteULike data set, additional 339,150 papers from DBLP (a superset of CiteULike) are included; and for news data set, additional 3,935,228 news articles are also included. 

The detailed data set statistics are shown in Table~\ref{tab:data_stat1} and~\ref{tab:data_stat2}.
\begin{table}[t!]
	\small
	\centering
	\caption{\label{tab:data_stat1} Data statistics for user, items and their interactions.}
	\begin{tabular}{lrrr}
		\toprule
		{} &  \# of user &  \# of item &  \# of interaction \\
		\midrule
		Citeulike         &       5,551 &      16,980 &            204,986 \\
		News             &      10,000 &      58,579 &            515,503 \\
		\bottomrule
	\end{tabular}
\end{table}
\begin{table}[t!]
	\small
	\centering
	\caption{\label{tab:data_stat2} Data statistics for text content.}
	\begin{tabular}{lrrrrr}
		\toprule
		{} &  voc. size & max & min & mean & median \\
		\midrule
		Citeulike (title)         &             4,777 &          15 &           2 &            9 &              9 \\
		Citeulike (title\&abs.)  &  23,011 &         300 &           22 &          194 &            186 \\
		News (title)         &            16,589 &          20 &           1 &           11 &             11 \\
		News (title\&sum.)    &            41,537 &         200 &           2 &           89 &             90 \\
		\bottomrule
	\end{tabular}
\end{table}

\subsection{Comparing Methods and Settings}

We compare following methods in experiments:

\begin{itemize}[leftmargin=2em]
	\item Cosine similarity matching~\cite{pazzani2007content}, which is based on similarities of TF-IDFs between candidate and user's historical items.
	\item Regularized multi-task logistic regression~\cite{evgeniou2004regularized}, which can be seen as one-layer linear text model.
	\item CDL (Collaborative Deep Learning)~\cite{wang2015collaborative}, which simultaneously trains auto-encoder for encoding text content, and matrix factorization for encoding user behavior.
	\item Content Pre-trained, which first pre-trains text embeddings by Paragraph Vector, and then used as fixed item features for matrix factorization.
	\item \modelsup{}. This is our proposed supervised framework. Note that two variants of text embedding function $f(x)$ are compared: MoV and CNN.
	\item \modelsemi{}. This is the joint text embedding framework. Both text embedding functions, MoV and CNN, are compared.
\end{itemize}

\textbf{Parameter Settings}: For CDL, both \modelsup{} and \modelsemi{}, we set the dimensions of both user embedding and final text embedding vector to 50 for fair comparisons. For CNN, we use 50 filters with filter size of 3. Regularization is added using both weight decay on user embedding and dropout on item embedding. We use Adam~\cite{kingma2015adam} with learning rate of 0.001. For both baselines and our model, we tune the parameters with grid search.

\textbf{Evaluation Metrics}: We adopt MAP (Mean Average Precision) and average AUC for evaluation. First, for each interaction between a user and a test item in the test set, we sample 10 negative samples from a test item-pool to form the candidate set. Then, AP and AUC are computed based on the rankings given by the model and the final MAP and average AUC are averaged over all users.

\begin{table*}[t!]
	\centering
	\begin{tabular}{cccccc}
		\toprule
		{} &                 CiteULike (title) &                   CiteULike (title\&abs.) &                       News (title) &               News (title\&sum.) \\
		\midrule
		Cosine   &  0.5535 / 0.8194 &                   0.7116 / 0.9162 &                    0.3526 / 0.6950 &            0.4580 / 0.7721 \\
		Multitask          &         0.6129 / 0.8441 &                    0.7355 / 0.9258 &                    0.4051 / 0.7085  &           0.4560 / 0.7760 \\
		Content Pretrained &  0.6250 / 0.8961 &                    0.7310 / 0.9372 &                    0.4512 / 0.8145 &           0.4778 / 0.8352 \\		CDL                &   0.6182 / 0.8839 &                   0.7484 / 0.9410  &                   0.3549 / 0.7648 &            0.4477 / 0.8060 \\
		\midrule
		\modelsup{} (MoV)   &     0.6789 / 0.9201  &                   0.7476 / 0.9432  &                   0.497 / 0.8294 &           0.5272 / 0.8515 \\
		\modelsup{} (CNN)   &     0.6908 / 0.9264 &                     0.7519 / 0.9458 &                    0.5069 / 0.8470 &           0.5227 / 0.8580 \\
		\modelsemi{} (MoV)     & \textbf{0.7073 / 0.9309}  &  \textbf{0.7641 / 0.9485} &                  0.5020 / 0.8462 &  0.5294 / \textbf{0.8628} \\
		\modelsemi{} (CNN)     &                   0.6990 / 0.9274 &         \textbf{0.7620 / 0.9478} &  \textbf{0.5149 / 0.8541} &  \textbf{0.5353 / 0.8626}
		 \\
		\bottomrule
	\end{tabular}
	\caption{\label{tab:result_main} Performance of different methods on four data sets. Two metrics are reported, namely MAP (left to slash), and AUC (right to slash). For both metrics, the larger the better.}
\end{table*}

\subsection{Performance Comparison}

Table~\ref{tab:result_main} shows MAP and AUC results of different methods on four data sets. As shown in the results, our methods (both TER and TER+) consistently beat other baselines and achieve state-of-the-art performance. Other several important observations can also be made from the results: 1) representation learning or embeddings methods (our methods, pre-trained method and CDL) can achieve better results compared to traditional TF-IDF based methods, 2) the joint supervised and unsupervised text embedding can achieve better results compared to supervised or unsupervised text embedding alone, and 3) the advantage of our model on short texts is more significant compared to longer one. We also observe that Mov outperforms CNN in some cases (e.g. in CiteULike data sets), we conjecture this is due to that words in CiteUlike may be more indicative w.r.t. user interests so simpler embedding functions can already well capture the semantics.

\begin{figure}[t!]
\small
\centering
\begin{subfigure}[b]{0.20\textwidth}
	\includegraphics[width=\textwidth]{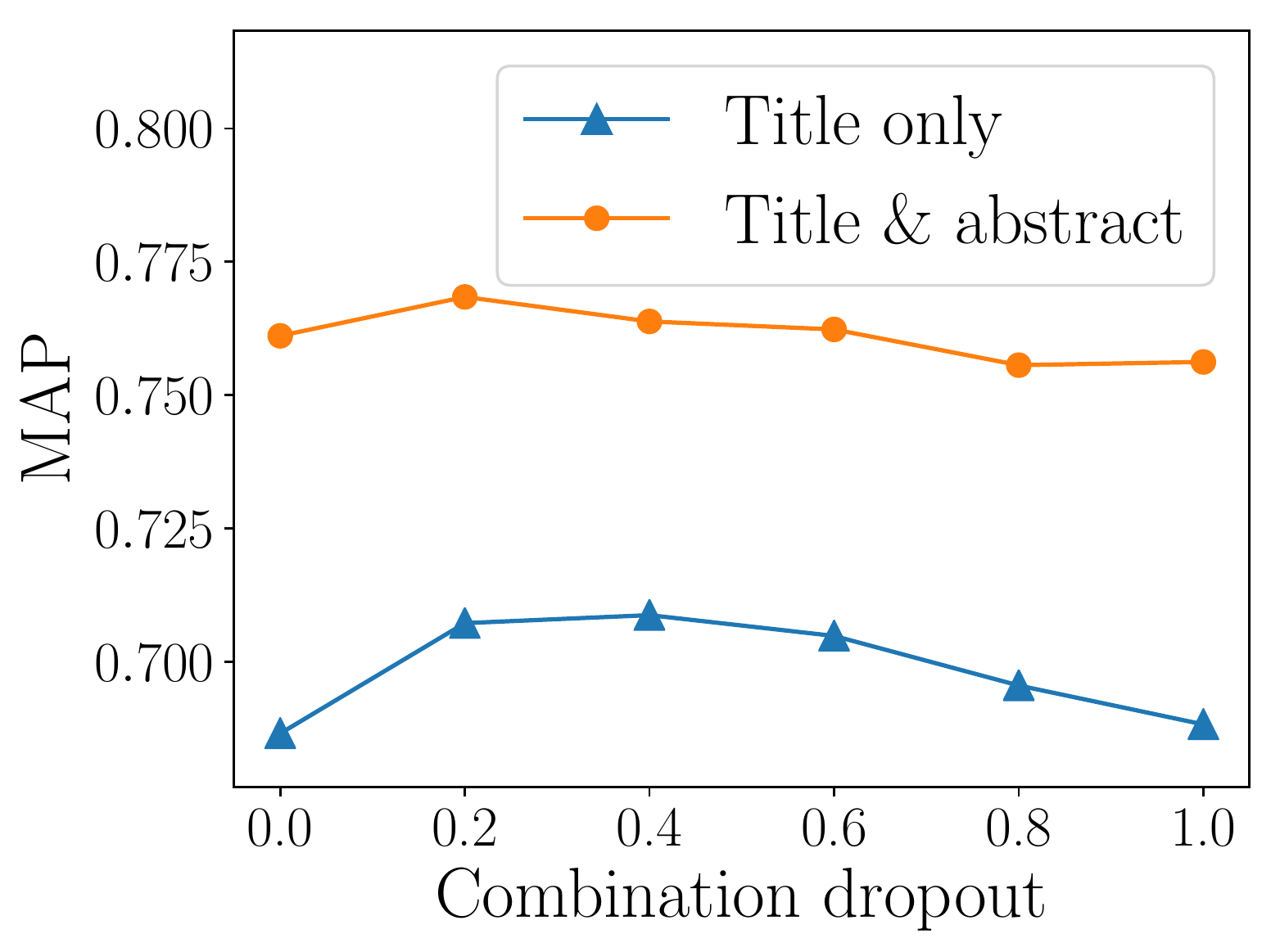}
	\caption{CiteULike}
	\label{fig:unsup_dropout_citeulike}
\end{subfigure}
\hspace{1em}
\begin{subfigure}[b]{0.20\textwidth}
	\includegraphics[width=\textwidth]{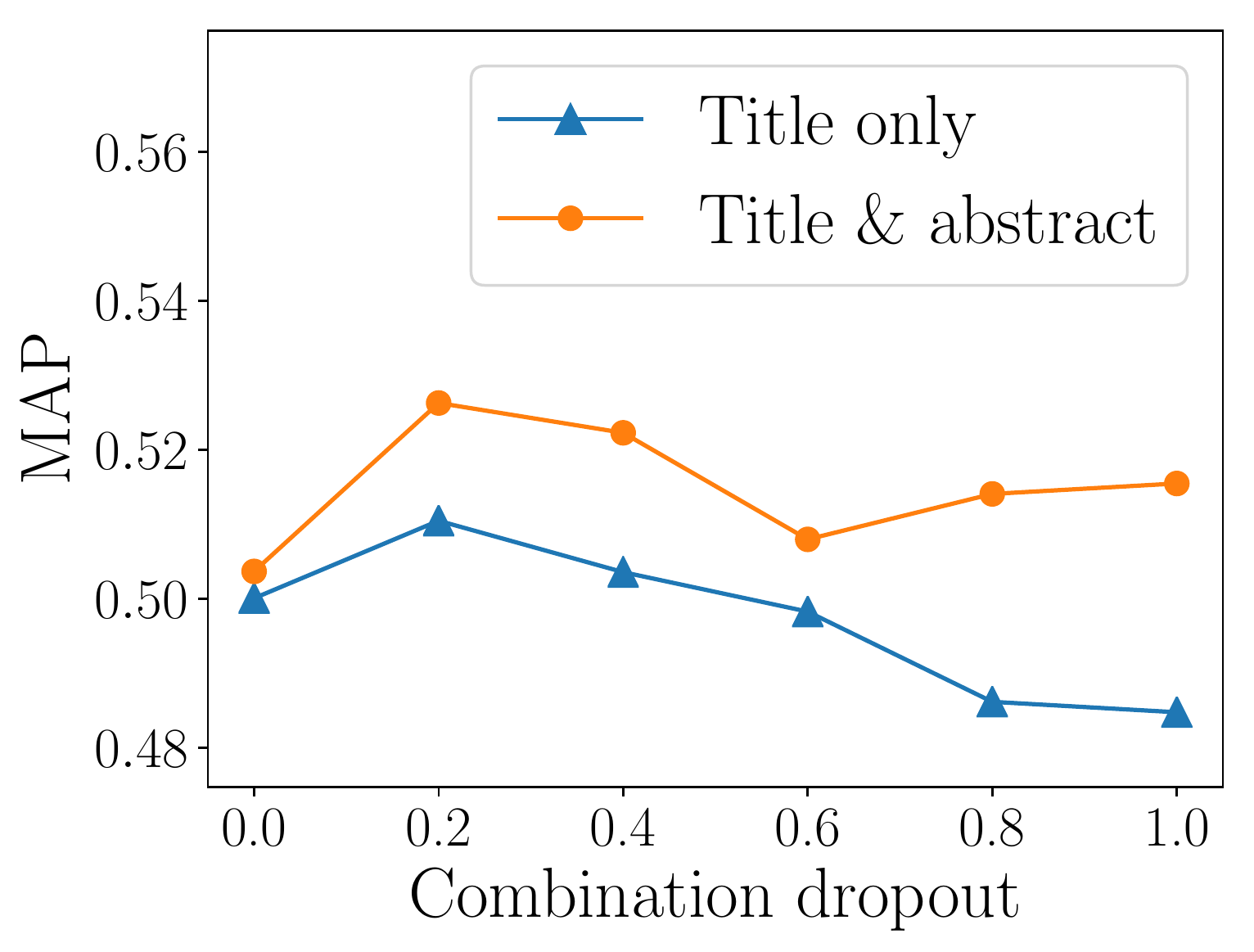}
	\caption{News}
	\label{fig:unsup_dropout_news}
\end{subfigure}
\caption{Mean Average Precision of different dropout rate for utilizing unsupervised text embedding.}\label{fig:unsup_dropout}
\end{figure}

Figure~\ref{fig:unsup_dropout} shows performances of different dropout rate for pre-trained text embedding vector $h_2$ in combination function $g(h_1, h_2)$. We observe that, as dropout rate increases, most of the curves go up and then go down. The peak occurs mostly around $0.2$ to $0.4$, both the $0$ and $1$ two extreme points have worse results. This further confirms the effectiveness of incorporating unsupervised text embedding, and also show that certain level of noise injected into pre-trained text embedding can improve performance.

\subsection{Case Studies}

\begin{figure*}[t!]
	\small
	\centering
	\begin{subfigure}[b]{0.2\textwidth}
		\includegraphics[width=\textwidth]{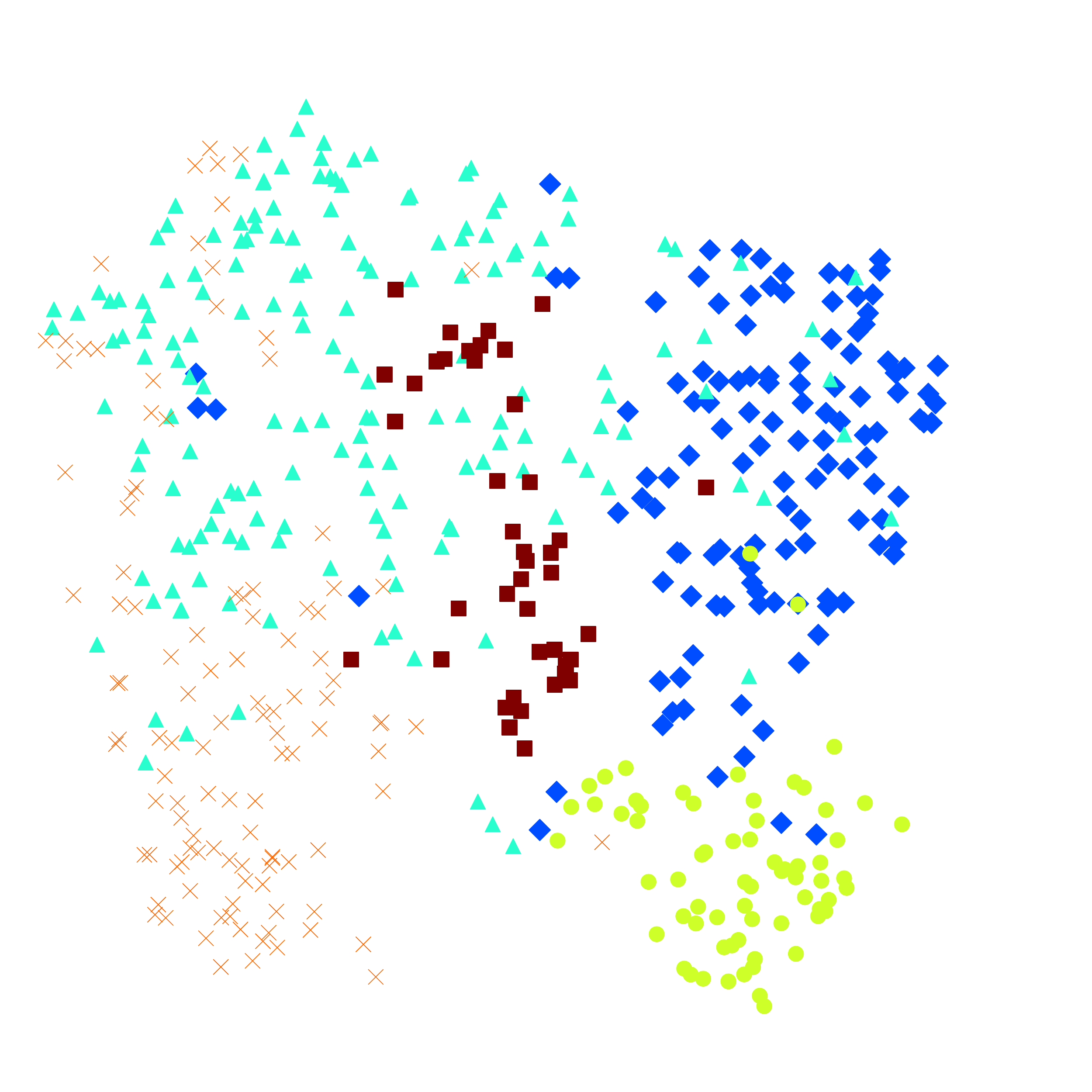}
		\caption{CDL}
		\label{fig:unsup_dropout_citeulike}
	\end{subfigure}
	\begin{subfigure}[b]{0.2\textwidth}
		\includegraphics[width=\textwidth]{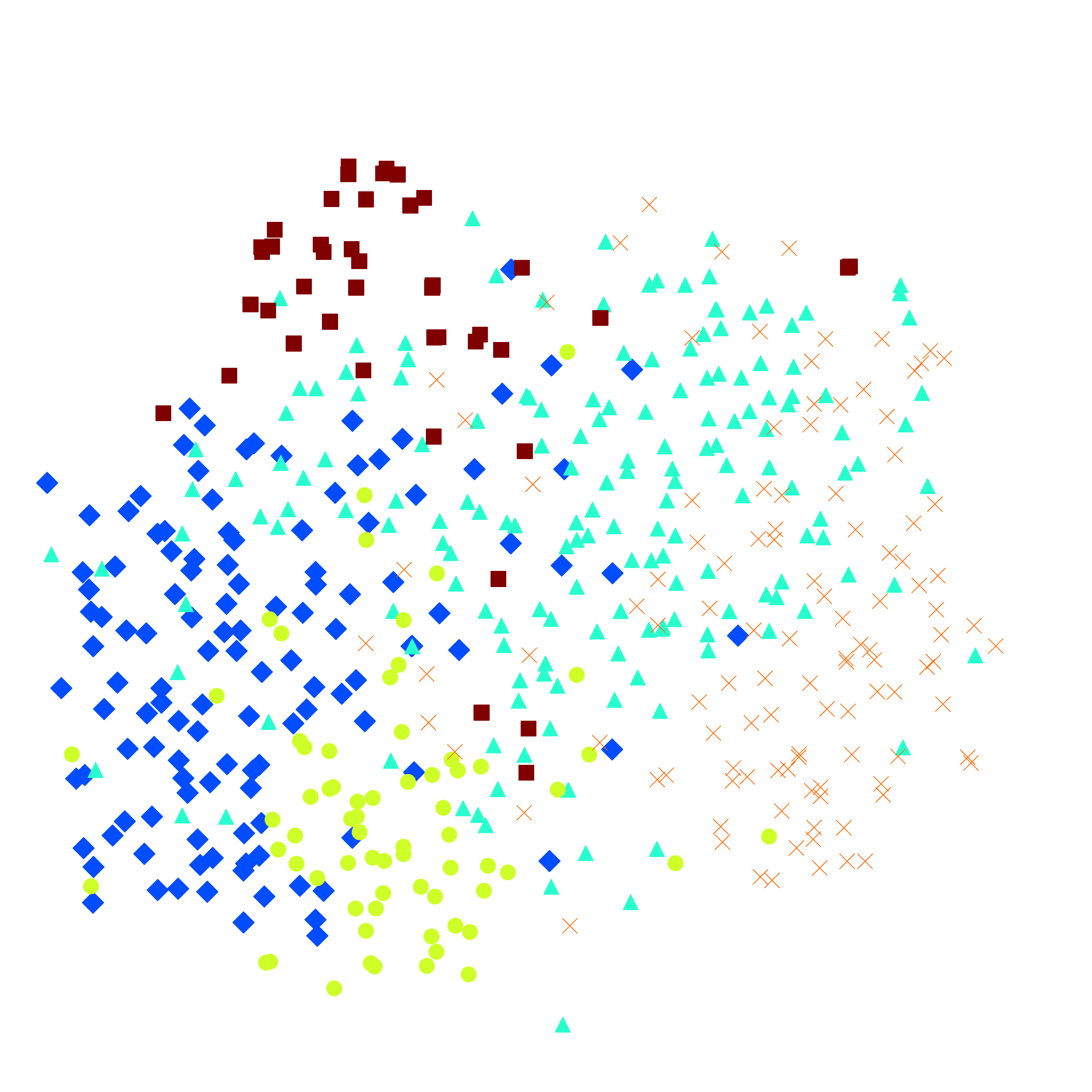}
		\caption{Pre-trained}
		\label{fig:unsup_dropout_news}
	\end{subfigure}
	\begin{subfigure}[b]{0.2\textwidth}
		\includegraphics[width=\textwidth]{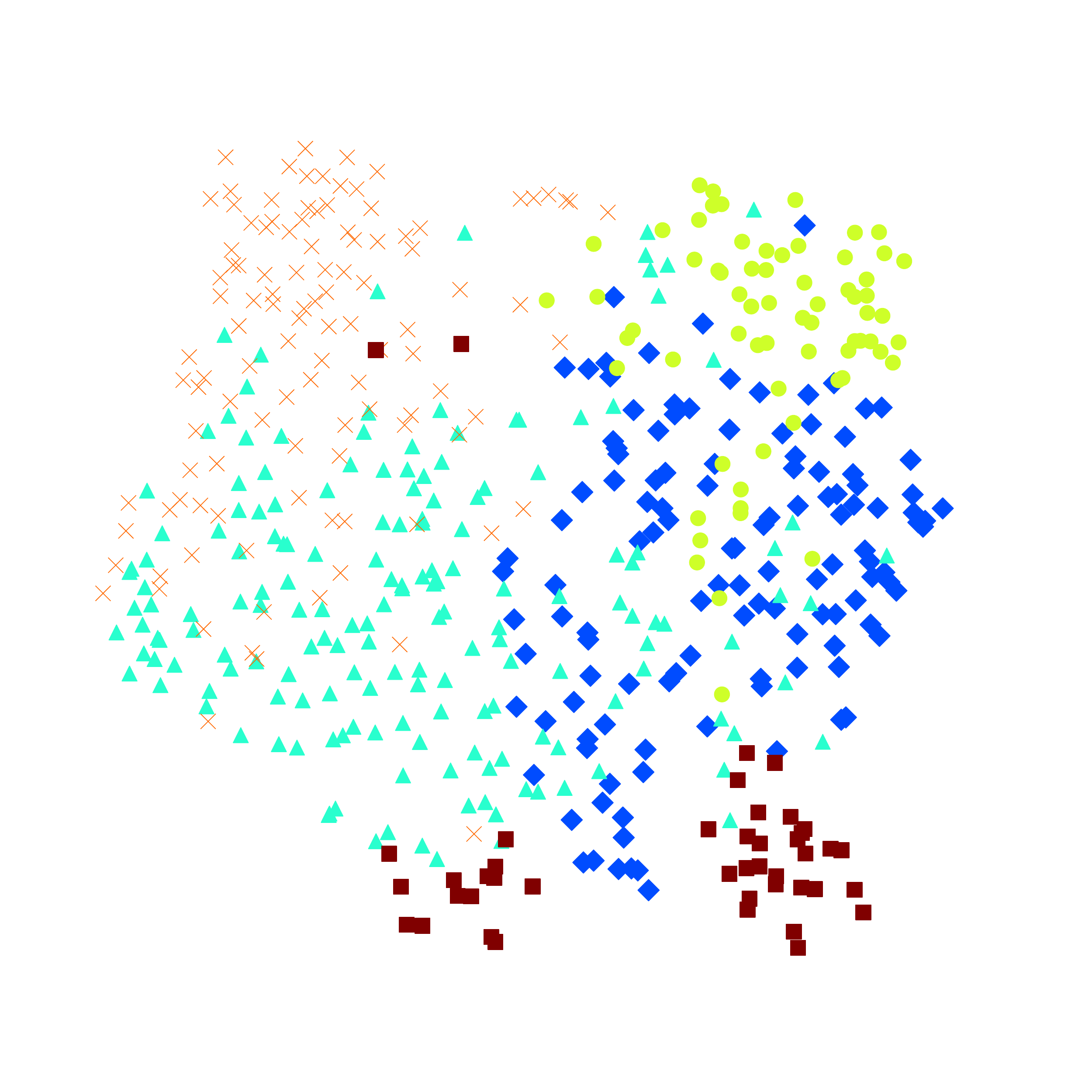}
		\caption{\modelsup{} (MoV)}
		\label{fig:unsup_dropout_news}
	\end{subfigure}
	\begin{subfigure}[b]{0.2\textwidth}
		\includegraphics[width=\textwidth]{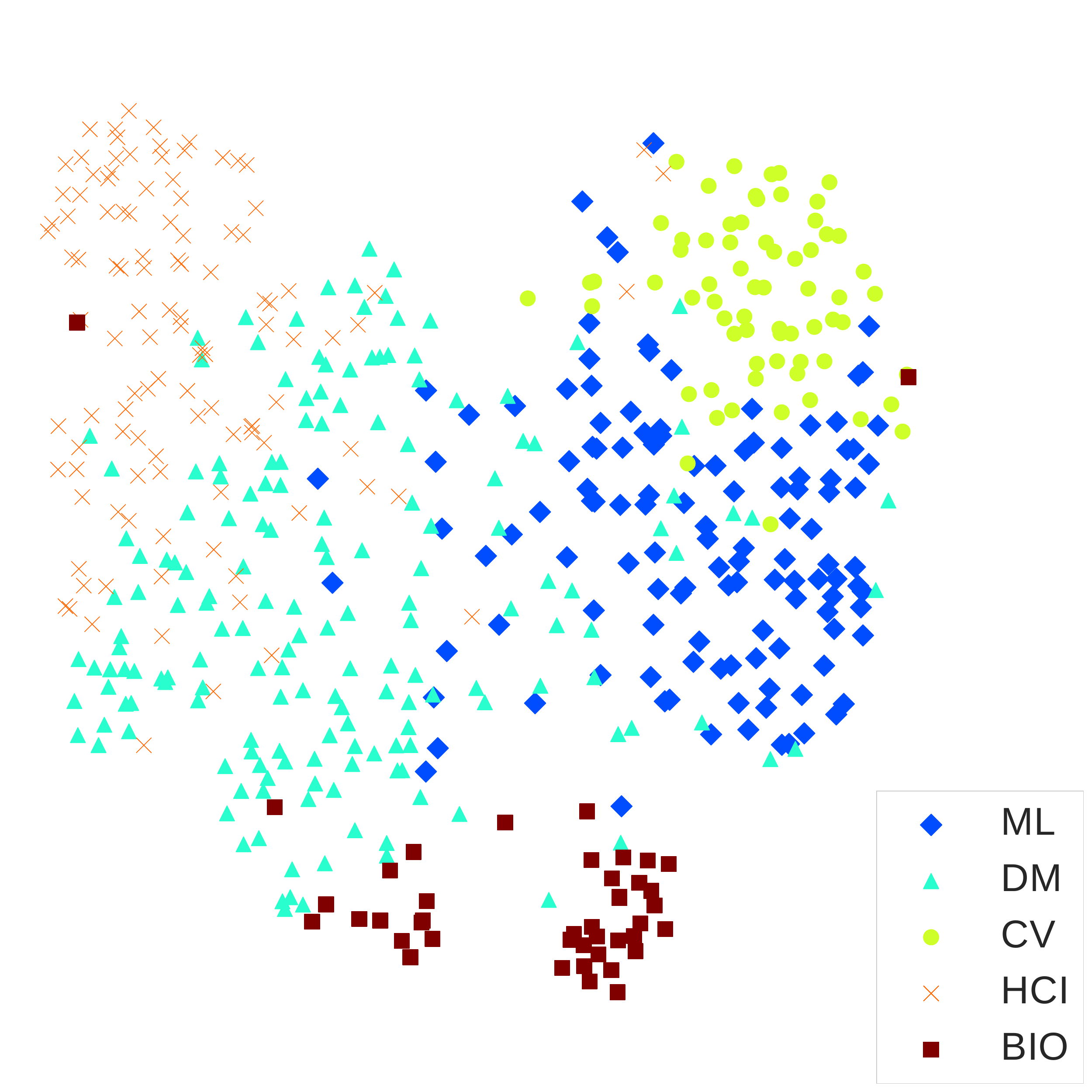}
		\caption{\modelsemi{} (MoV)}
		\label{fig:unsup_dropout_news}
	\end{subfigure}
	\caption{Article embeddings for papers of different domains.}\label{fig:article_emb}
\end{figure*}

\begin{table*}[!t]
	\small
	\centering
	\begin{subtable} {1\linewidth}
		\centering
		\begin{tabular}{lp{10cm}}
			\toprule
			{Unsupervised} &  recurrent feedforward artificial feed multilayer trained neuron chaotic parameters \\
			{Supervised} & cortex motor spike hippocampal attention sensory train parietal perceptual \\
			\bottomrule
		\end{tabular}
		\caption{\label{tab:case11} Most similar words to ``neural".}
	\end{subtable}
	\begin{subtable} {1\linewidth}
		\centering
		\begin{tabular}{lp{10cm}}
			\toprule
			{Unsupervised} &  training styles learners experts contexts activities reinforcement traditional concept 
			\\
			{Supervised} & measuring rehabilitation courses multimodal elearning special review sense instruction \\
			\bottomrule
		\end{tabular}
		\caption{\label{tab:case12} Most similar words to ``learning".}
	\end{subtable}
	\caption{\label{tab:case1} Most similar words according to the learned word embedding.}
\end{table*}

\begin{table*}[t!]
	\small
	\begin{subtable}{1\linewidth}
		\centering
		\begin{tabular}{lp{7.7cm}p{7.7cm}}
			\toprule
			{} &                                    Pretrained &  \modelsemi{} (MoV)    \\
			\midrule
			1  &                                Portraits of complex networks &                Navigability of complex networks \\
			2  &      \textit{Exploring the diversity of complex metabolic networks} &                   Portraits of complex networks \\
			3  &                          Synchronization in complex networks &   Hierarchical organization in complex networks \\
			4  &                 \textit{Lethality and centrality in protein networks} &                             Scale free networks \\
			5  &   \textit{Exploring the new world of the genome with dna microarrays} &         \textit{Structure of growing social networks} \\
			\midrule
		\end{tabular}
		\caption{\label{tab:case22} Queried paper: Exploring complex networks.}
	\end{subtable}
	\begin{subtable} {1\linewidth}
		\centering
		\begin{tabular}{lp{7.7cm}p{7.7cm}}
			\toprule
			{} &                                    Pretrained &                                           Combined \\
			\midrule
			1  &                       Communities in networks &               Semiotic dynamics in online social communities \\
			2  &             \textit{Scientific computing in the cloud} &                                    Communities in cyberspace \\
			3  &                  Communities and technologies &   Audience , structure and authority in the weblog community \\
			4  &                                \textit{Cloud computing} &                                 Communities and technologies \\
			5  &                       \textit{Computing with membranes} &  Conversations in the blogosphere an analysis from the bo... \\
			\bottomrule
		\end{tabular}
		\caption{\label{tab:case21} Queried paper: Mapping weblog communities.}
	\end{subtable}
	\caption{\label{tab:case2} Most similar articles found based on different text embedding. We use \textit{italics} to highlight those ``inaccurate" results.}
\end{table*}

To further understand the proposed model, we conduct several case studies looking into the layout or nearest neighbors of words and articles in the embedding space. 

To visualize the text embedding learned from different models, we firstly choose top conferences in five domains (ML, DM, CV, HCI, BIO), and then randomly select articles that are published in those conferences. We apply TSNE \cite{maaten2008visualizing} to visualize 2d map for these articles, and color them according to their domains of publication. The results are shown in Figure~\ref{fig:article_emb} where we found that our combined model can best distinguish papers from different domains.

Table~\ref{tab:case1} shows similar words for given queried words, i.e. ``neural" and ``learning", in CiteULike data set. From the result we clearly see the distinction between meanings of word learned from both methods. For example, the nearest word ``neural" learned in unsupervised text embedding (articles with and without user like behavior) is mostly related to artificial neural networks, but in supervised text embedding, it is mostly related to neuroscience, which is more close to biology. This is because that in the CiteULike data set, there exist a lot of biologists, so the word embedding learned from supervised text embedding is likely to be dominated by the neuroscience perspective. However, by incorporating the unsupervised text embedding learned from a larger corpus, more meanings of the words can be recovered.

Table~\ref{tab:case2} shows the similar articles given a randomly selected queried article. We find that although unsupervised text embedding can provide some similar articles, the proposed framework (both \modelsup{} and \modelsemi{}) can better capture the similarity of articles.

%% file: related.tex
\section{Related Work}

Our work is related to both personalized recommendation and text embedding and understanding.

Collaborative filtering \cite{koren2009matrix} has been one of the most effective methods in recommender system. Methods like matrix factorization \cite{koren2008factorization,salakhutdinov2011probabilistic} are wide adopted, and recently some methods based on neural networks are also explored \cite{wang2015collaborative,sedhain2015autorec,zheng2016neural}. Content based methods are proposed \cite{pazzani2007content,chen2017task}, but has not been well developed to exploit deep semantics of content information. Hybrid methods can improve so-called ``cold-start" issue by incorporating side information \cite{rendle2010factorization,chen2012svdfeature,singh2008relational}, or item content information \cite{wang2011collaborative,gopalan2014content,wang2015collaborative}. In our case, although we have historical data about users' interactions with items, but at the time of recommendation we are considering the items that have never been seen before, which cannot be handle directly by most existing matrix factorization based methods. Our model is similar to CDL \cite{wang2015collaborative}, but with following differences: (1) we treat the problem as ranking instead of rating prediction problem, (2) we provide a general framework which allows flexible choice of text embedding function $f(x)$, and (3) our model can explicitly incorporate unsupervised text embedding. 
 
To understand text data, both supervised and unsupervised methods are proposed. Supervised methods are usually guided by text labels, such as sentiment labels or category labels. Different from traditional text classification, which train SVM or logistic regression classifiers based on n-gram features \cite{joachims1998text,pang2002thumbs}, recent work take advantage of distributed representation brought by embedding methods, which include CNN \cite{collobert2011natural,kim2014convolutional,zhang2015character}, RNN \cite{tang2015document} and others \cite{joulin2016bag}. Those methods cannot be directly applied for recommendation as only a global classification/ranking model is provided. Also, instead of using labels as in existing supervised text embedding methods, we utilize user item interactions as supervision to learn the text embedding function. There are also unsupervised text embedding techniques \cite{mikolov2013efficient,mikolov2013distributed,le2014distributed}, which do not require labels but cannot adapt to the task of interest.

We further generalize the proposed model and develop efficient training techniques in \cite{chen2017onsampling}.

%% file: conclusion.tex
\section{Conclusions}

In this work, we tackle the problem of content-based recommendation for completely new texts. An novel joint text embedding based framework is proposed, in which user embedding and text embedding function are learned end-to-end based on interactions between users and items. The text embedding function is flexible, and can be specified by deep neural networks. Both supervised and unsupervised text embeddings are fused together by an combination module as part of a unified model. Empirical evaluations based on real-world data sets demonstrate that our model can achieve state-of-the-art results for recommending new texts. As for the future work, it is interesting to explore other ways of incorporating unsupervised text embeddings.